\documentclass{article}

\usepackage[english]{babel}

\usepackage[letterpaper,top=2cm,bottom=2cm,left=3cm,right=3cm,marginparwidth=1.75cm]{geometry}

\usepackage{amsmath}
\usepackage{graphicx}
\usepackage[colorlinks=true, allcolors=blue]{hyperref}

\title{A more inclusive effective dark fluid equation of state parameter: constraints from SKA and Euclid like surveys}
\author{Ziad Sakr  \\ziad.sakr@ift.csic.es\\\\
\normalsize{Instituto de Física Teórica UAM-CSIC, Campus de Cantoblanco, 28049 Madrid, Spain}\\
\normalsize{IRAP, Universit\'e de Toulouse, CNRS, CNES, UPS, Toulouse, France} \\
\normalsize{Faculty of Sciences, Universit\'e St Joseph; Beirut, Lebanon}
 }
\date{}

\begin{document}
\maketitle

\begin{abstract}
We forecast constraints on an effective dark fluid equation of state parameter $w_{\rm eff}$ that encapsulates modified gravity theories that modifies both the Universe background expansion as well as its large scale structures growth. This is achieved through relating Friedmann equations' dark fluid pressure and density content, thus $w_{\rm eff}$, to modified gravity parameterized models by mean of the Newtonian potential equation parameter $\mu_0$, the gravitational slip parameter $\eta_0$ and a redshift dependent Hubble parameter $H_{0,{\rm bck}}$. We adopt next stage SKA survey specifications, alone or in combination with concurrently expected DR3 Euclid survey release, paying attention to the modeling and recipe of the implementation of the galaxy clustering and lensing probes obtained from the two surveys. We consider two data mock models: one with deviation of the intermediate parameters at the level of 10 \% (yielding however $w_{\rm eff}=-1.03$) and another sub-percently close to $\Lambda$CDM. We found that the three parameters deviation from $\Lambda$CDM could only be detected at 1~$\sigma$ from SKA alone, while this improves to $\sim$ 2~$\sigma$ when we combine with Euclid. An improvement of the order of 30\% on the bounds is reached after projecting the three parameters into a single $w_{\rm eff}$ parameter. However, this affects both cases and thus it does not change much, though it improves, the level of detection with respect to $\Lambda$CDM values. We conclude that synergy from both surveys benefits to tighten our constraints, but also that our highly generalized parameterization with \textit{percent deviations} from the fiducial model, although impacting at both the background and the perturbation level, will be hard to disentangle from $\Lambda$CDM at the level at which our forecast is performed and it still needs, to the least, data from more advanced stages of the adopted surveys to hope reach this target.
\end{abstract}

\section{Introduction}\label{sect:intro}

The latest results by Stage-IV DESI survey \cite{DESI:2024mwx}, showing a preference for a phenomenological parameterized dynamical dark energy equation of state over a simple cosmological constant cold dark matter model governing the expansion of the Universe, has sparkled interest in such models. A possible confirmation from next generation surveys could clearly signify a change of paradigm in the Universe's adopted concordance model.

Though the Chevallier–Polarski–Linder (CPL) parameterization \cite{Chevallier:2000qy,Linder:2002et} is the commonly used one to detect the possibility of a dynamical dark energy (see however e.g. \cite{Sakr:2025daj} or \cite{Blanco:2025vva} for a critic of this parameterization). There has been since then many propositions for other more or less general phenomenological approaches to explore in the light of the new data, such as \cite{Li:2025ops,Lee:2025pzo,Blanco:2025vva,Shajib:2025tpd,DESI:2025wyn,Duchaniya:2025oeh}

Here we propose a new general phenomenological, yet within a theoretical framework, model for an effective dark fluid equation of state $w_{\rm eff}$ (see \cite{Nesseris:2022hhc} for a review of effective fluid approach and references there in for different models within it) that impacts at both the universe background expansion and perturbations level and elaborate on the role that the SKA will play in synergy with other forthcoming deep survey Euclid in constraining or excluding it with respect to a $\Lambda$CDM  equivalent value.

The manuscript is structured in the following: after introducing the subject of study here in section~\ref{sect:intro}, we lay down the theoretical foundation in section~\ref{sect:theoframe}, followed by a description of the data and the modeling of the observables in section~\ref{sect:datamod}, then we show, discuss and conclude on the forecast outcomes in section~\ref{sect:forecasts}. 

\section{Theoretical framework}\label{sect:theoframe}

We start with the following 4-dimensional action where $ f $ is a function of the Ricci scalar $ R $, a scalar field $\phi$, and a kinetic term $ X $, that one can specify to a class of theories where the gravity Lagrangian is given by
\begin{equation}
\mathcal{L}=\frac{F(\phi)}{2}R+X-U(\phi)\, , \label{eq:SCTaction}
\end{equation}
where $U(\phi)$\footnote{that ends up by being chosen as equal to 2$\Lambda$} is the potential of the scalar field and $F(\phi)$ a function determining its coupling to gravity

\begin{equation}
S = \int d x \sqrt{-g} \left( \frac{1}{2} \, \frac{1} {f(R, \phi, X) + \mathcal{L}_m}  \right).
\end{equation}
which allows us to write a generalization of the Friedmann equations on the background side,
\begin{equation}\label{eq:MF1}
3FH^2 = f_X X + (FR - f) - 3H\dot{F} + \rho_m
\end{equation}
\begin{equation}\label{eq:MF2}
-2F\dot{H} = f_X X + \ddot{F} - H\dot{F} + \rho_m
\end{equation}
where $f_{,X}=\partial_Xf(R,\phi,X)$;
while the perturbation side is extended through the modified gravity potentials,
\begin{align}
 -k^2\,\Psi(a,\vec{k}) &= \frac{4\pi\,G}{c^2}\,a^2\,\bar\rho(a)\,\Delta(a,\vec{k})\times\mu(a,\vec{k}), \label{eq:mu}\\ 
 -k^2\,\left[\Phi(a,\vec{k})+\Psi(a,\vec{k})\right] &= \frac{8\pi\,G}{c^2}\,a^2\,\bar\rho(a)\,\Delta(a,\vec{k})\times\Sigma(a,\vec{k}), \label{eq:sigma}
\end{align}
where we follow the approach of \cite{Tsujikawa:2007gd} that translates the action terms into the modified gravity functions:
 \begin{align}
\mu(a,k) &= \frac{1}{F(\phi)}\frac{F(\phi)+2F_{,\phi}^2}{F(\phi)+\frac32F_{,\phi}^2}\, ,\label{eq:mueff}\\
\eta(a,k) &= \frac{F_{,\phi}^2}{F(\phi)+2F_{,\phi}^2}\, ,\label{eq:etaeff}
\end{align}
where $F_{,\phi}=\partial_\phi F(\phi)$, yielding in quasi-static slow varying scalar field assumption \cite{ Sakr:2021ylx}
\begin{align} 
 F &= \frac{2}{\mu + \mu\eta}\, ,\label{eq:eff_1}\\ 
 \dot F &= - \frac{2(\dot \mu (1+\eta)+ \mu \dot\eta)}{(\mu + \mu\eta)^2}\, ,\label{eq:eff_2}\\
 \ddot F &= 4 \frac{(\dot \mu(1+\eta) + \mu \dot \eta)^2}{(\mu + \mu\eta)^3} - 2 \, \frac{2 \,\dot \mu \dot \eta + \dddot \mu (1+\eta)+\mu \ddot \eta}{(\mu + \mu\eta)^2}\, \label{eq:eff_3}\, .
\end{align}
Now we can rewrite Eq.~\ref{eq:MF1} and \ref{eq:MF2} as follows:
\begin{align}
 3F_0 H^2 & =\rho_{DF}+\rho_m,\\
-2F_0 \dot{H} & =\rho_{DF}+p_{DF}+\rho_m,
\end{align}
allowing us to compute an effective dark fluid equation of state parameter:
\begin{equation}
w_{\text{eff}} = \frac{p_{DF}}{\rho_{DF}} = -1 + \frac{2\ddot{F}  - 4H\dot{F} - 4\dot{H} (F_0 - F)}{FR - f - 6H \dot{F}+ 6H^2 (F_0 - F)}
\end{equation}
where $F_0 = F(a=1)$ and we consider the following late-time parameterization for $\mu$ and $\eta$, but also, to generalize further, we apply it as well to the Hubble parameter (function) :
\begin{equation}
\mu = 1 + \mu_0 \, a, \quad \eta = 1 + \eta_0 \, a, \quad H_{\rm eff} = H (1 + H_{0, \text{bck}} \, a)
\end{equation}

\section{Datasets and the modeling of observables}\label{sect:datamod}

We use and modify the cosmological solver \texttt{MGCLASS} \cite{Sakr:2021ylx} to incorporate
our models and parameterisations, and \texttt{CosmicFish} \cite{Casas:2022vik,Sakr:2025kgq} to integrate them into
the considered probes after modifying the recipes (for more, see detailed $\Lambda$CDM equations taken from \cite{Casas:2022vik} and \cite{Euclid:2025tpw}, when necessary to account
for our extensions, in particular the lensing prefactor is modified  because the conservation Friedmann equation is changed in our model to become
\begin{equation}
\frac{\dot{\rho_m}}{\rho_m} = H (1 + H_{0, \text{bck}} a)
\end{equation}
leading to, after solving the differential equation 
\begin{equation}
\rho_m = \rho_0 \frac{a^{-3}}{e^{3(a-1)}}
\end{equation}
which modifies the prefactor into:
\begin{equation}
\frac{3}{2} (1+z) H_0 \, \Omega_{m,0} \quad -> \quad \frac{3}{2} (1+z) H_0 \, \Omega_{m,0} \frac{1}{e^{ -3 z/(1+z)} } 
\end{equation}
while the eNLA intrinsic alignment is modified following the approach of \cite{Sakr:2025kgq} where $\mu$ following our model in this case is integrated in the Poisson potential equation governing attraction between point sources. For GCsp, we adopt a semi-analytic nonlinear RSD model, and cut at $k_{\rm max} \sim 0.15 \, h$ Mpc$^{-1}$ with free terms for each bin, that also includes Fingers-of-God and BAO damping \cite{Euclid:2025tpw}. We combine the 3x2pt Galaxy Clustering and Lensing probe from the radio Continuum survey with HI Spectroscopic Galaxy Clustering and Intensity Mapping data from AA* SKA design after rescaling the survey Redbook specifications \cite{SKA:2018ckk} accordingly. We additionally consider 3x2pt Photometric data combined with Spectroscopic Galaxy Clustering data from Euclid-like similar to forthcoming DR3 release \cite{Euclid:2024yrr} expected at the same time as our SKA primary probes. We stay at the linear level and cut at the conservative value (see \cite{Euclid:2019clj}) of $\ell \sim 750$. The parameters fiducials and specifications used are presented in Table~\ref{tab:baseline_parameters} and Table ~\ref{tab:Surveys_spec} but we emphasize that these are only to determine the sensitivity of alike surveys to our model and we are not providing official specifications for the actual present or future experiments.

\begin{table*}
\centering
\resizebox{0.9\textwidth}{!}{
\normalsize
\begin{tabular}{l|ccccccccccc}  
\hline \\
\text{Cosmological Parameters} & $\Omega_{\rm b,0}$ (fixed) & $n_s$ & $\Omega_{\rm m,0}$ & $h$ & $\sigma_8$ \\ \\

 & 0.05 & 0.96 & 0.32 & 0.67 & 0.815 \\ 
 
\hline \\

\text{Nuisance parameters} & $\mathcal A_{\rm IA}$ & $\eta_{\rm IA}$ & $\beta_{\rm IA}$ & $\mathcal{C}_{\rm IA}$ (fixed) & $\sigma_{\rm p}(z_1)$ & $\sigma_{\rm p}(z_2)$ & $\sigma_{\rm p}(z_3)$ & $\sigma_{\rm p}(z_4)$ & $\sigma_{\rm p}(z_5)$ & $\sigma_{\rm p}(z_6)$ \\ \\

& 1.72 & -0.41 & 2.17 & 0.0134 & 4.484 & 4.325 & 4.121 & 3.902 & 3.683 & 3.475 \\
\hline
\end{tabular}}
\caption{Cosmological and nuisance parameter adopted fiducial values. The bias $b_i$ for Euclid-like are obtained as $\sqrt{1+z_i}$ while SKA-IM are following $b_i =0.3 (1 + z_i) + 0.6$, and SKA-HI following table A8 of \cite{Casas:2022vik}. Finally, each bin has an additional nuisance shooting free parameter  $P_{{\rm shot},i}$ with fiducials equal to zero.}
\label{tab:baseline_parameters}
\end{table*}

\renewcommand{\arraystretch}{1.7}
\begin{table*}
\centering
\resizebox{\textwidth}{!}{
{
{\LARGE
\begin{tabular}{l|ccccccc} 
\hline
\text{Survey/Parameter} & $A_{\rm surv}({\rm deg}^2)$ & $z_{{\rm obs},i} ({\rm edges})$ & $\bar n_{\rm gal}(\mathrm{arcmin}^{-2})$ & $\sigma_\epsilon$ & $\ell_{\rm min}$ & $\ell_{\rm max}(z)$ \\
\hline
\textbf{Euclid photo  3$\times$2pt }  & 5000 & $\{0.001, 0.2, 0.39, 0.5, 0.65, 0.83, 1.03, 2.50\}$ & 35 & 1.5 & 10 & $750$ \\ 
\textbf{SKA Cont.  3$\times$2pt }  & 5000 & $\{0.001, 0.418, 0.560, 0.678, 0.789, 0.900, 1.019, 1.155, 1.324, 1.576, 2.50\}$ & 2.43 & 1.25 & 10 & $750$ \\
\hline
\hline
\text{Survey/Parameter} & $z_{{\rm obs},i} ({\rm edges})$ & $ dN/d\Omega dz \, (deg^{-2})$ & $V_i \, (h^{-3} {\rm Gpc}^{3})$ & $\sigma_{0,z}$ & $k_{\rm min}(h\,{\rm Mpc}^{-1})$ & $k_{\rm max}(h\,{\rm Mpc}^{-1})$ \\
\hline
\textbf{Euclid spectro} &  $\{0.9, 1.1, 1.3, 1.5, 1.8\}$ & $\{1815.0, \, 1701.5, \, 1410.0, \, 940.97\} $ & $\{2.68,  \,3.096,  \,3.40,  \,5.49 \}$ & 0.002 & 0.005 & 0.15 \\
\textbf{SKA HI} &  $\{0.0, 0.1, 0.2, 0.3, 0.4, 0.5\}$ & $\{1357.41, \, 1361.185, \, 632.0, \, 233.5215, \, 77.1765,\} $ & $\{0.013,  \,0.083,  \,0.201,  \,0.35,  \,0.52 \}$ & 0.002 & 0.005 & 0.15 \\ 
\hline
\hline
\text{Survey/Parameter} & $z_{{\rm obs},i} ({\rm edges})$ & $ \Omega_{\rm HI}(z_i)$ & $V_i \, (h^{-3} {\rm Gpc}^{3})$ & $\sigma_{0,z}$ & $k_{\rm min}(h\,{\rm Mpc}^{-1})$ & $k_{\rm max}(h\,{\rm Mpc}^{-1})$ \\
& $D_{\rm dish} \rm (m) $ & \text{Total Time} \rm (hours) & $N_{\rm dish}$ \\
\hline
\textbf{SKA IM} &  $\{0.5, 0.7, 0.9, 1.1, 1.3, 1.5, 1.7, 1.9, 2.1, 2.4, 2.6\}$ & $ \sim 4.0 (1+z_i)^0.6 \times 10^{-4}$ & $ = \frac{\Omega}{3}\, [r^3(z_{i+1})-r^3({z_i})] $ & 0.001 & 0.005 & 0.15 \\  
&  $15 $ & $10000 $ & $144$  \\  
\hline
\hline
\end{tabular}
}
}
}
\caption{Euclid-like DR3 expected photometric angular lensing and clustering, and spectroscopic survey 3D power spectrum specifications, with $A_{\rm surv}$ the survey area, $V_i$ the survey volume in each redshift bin, $\sigma_\epsilon$ the intrinsic ellipticity dispersion, and $\sigma_{0,z}$ the error on the photometric redshift measurement. Showing also specifications for SKA AA* configuration for the continuum angular lensing and clustering survey, the HI  and the Intensity Mapping MI power spectrum used in our forecast with $D$ the diameter of the dish, $N$ their number along with the number of observed hours, $r$ the comoving distance and $\Omega$ the solid angle of the volume covering region between two redshifts.}
\label{tab:Surveys_spec}
\end{table*}

\section{Forecast outcomes and conclusions}\label{sect:forecasts}

\begin{figure}
\centering
\includegraphics[width=0.49\columnwidth]{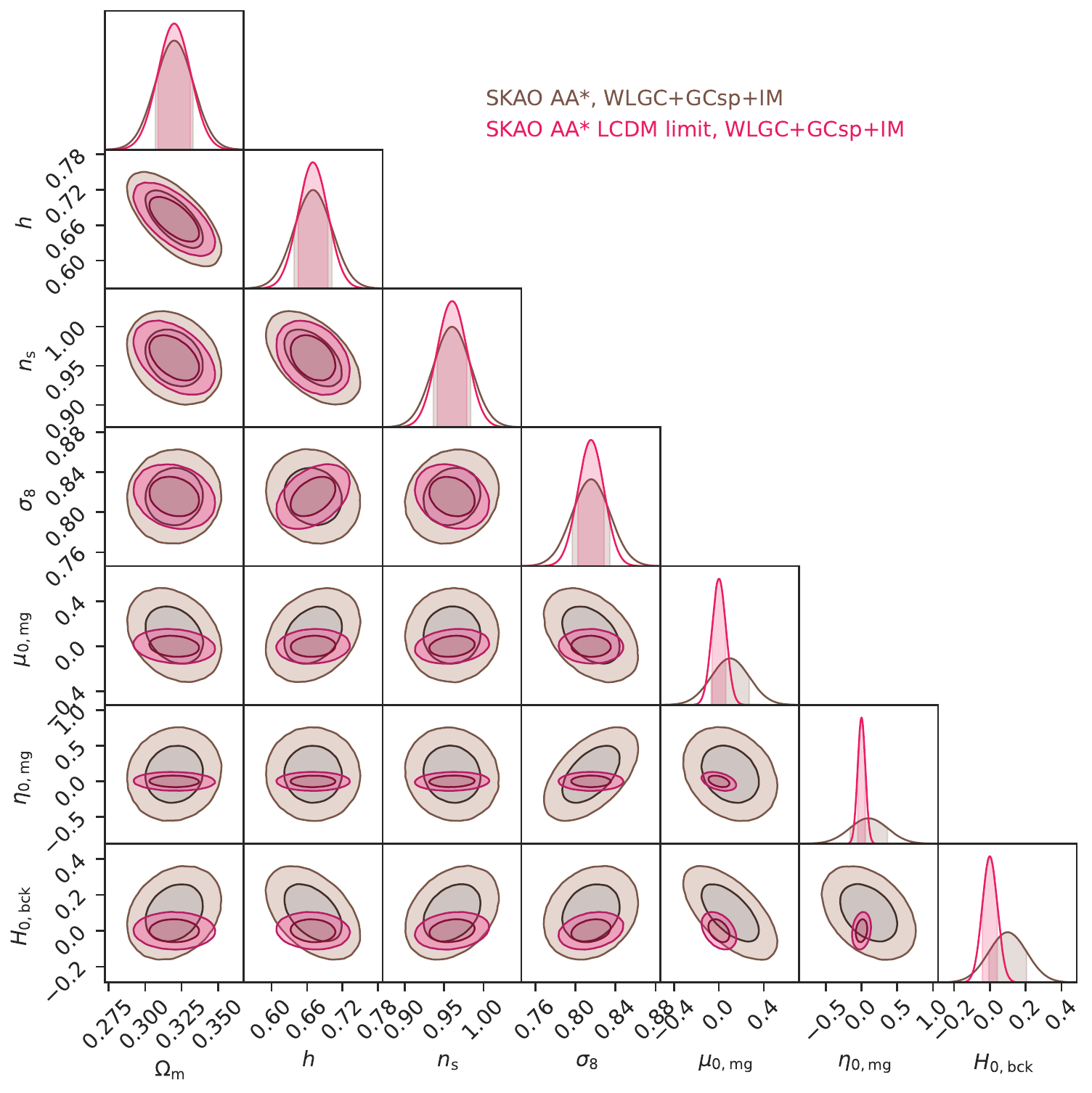}
\includegraphics[width=0.49\columnwidth]{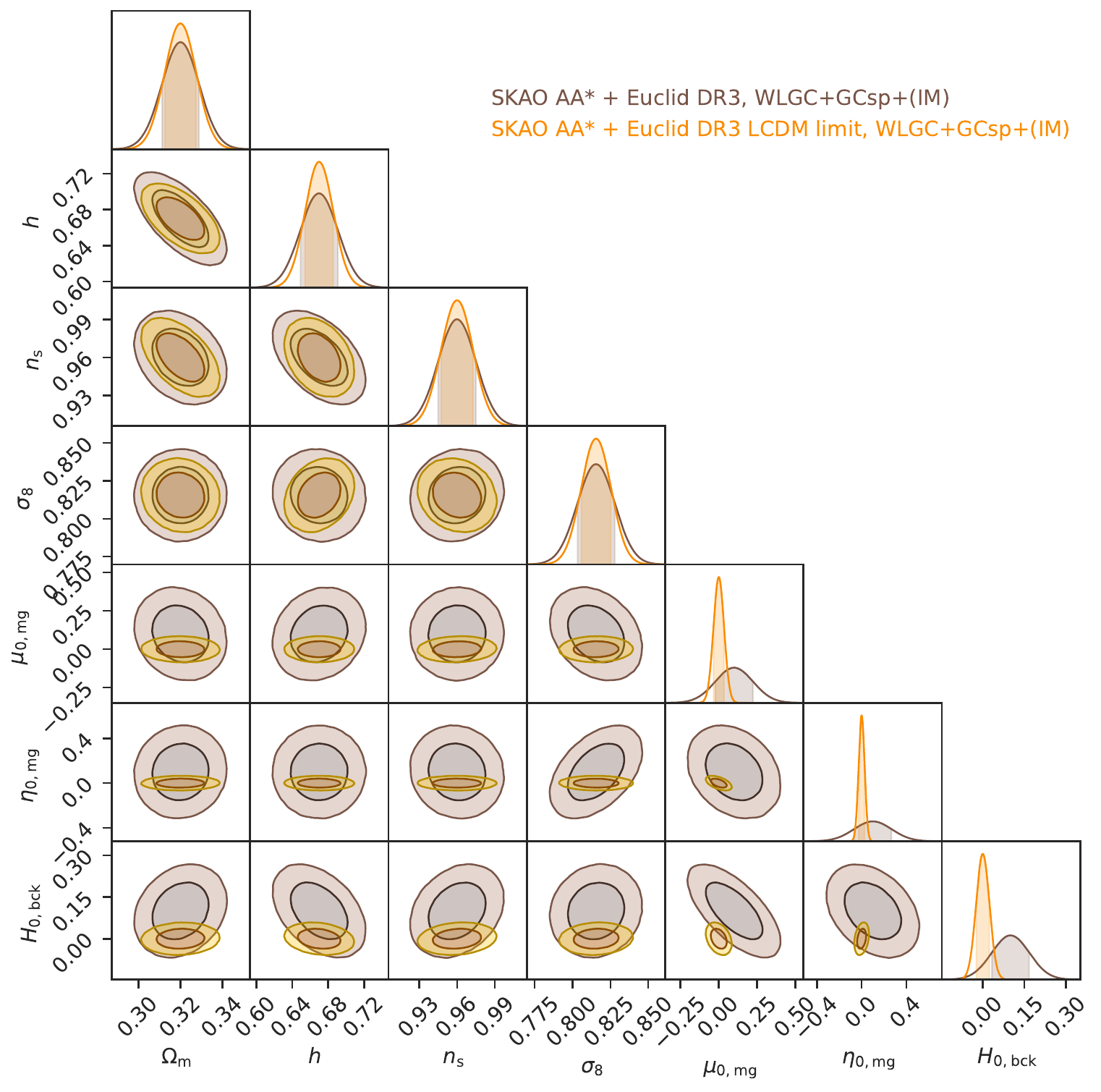}
\caption{Left plot: Forecast 1- and 2-$\sigma$ contours on the cosmological and model extensions parameters using SKAO continuum, HI probes and their combination following the AA* specifications for the far and close to $\Lambda$CDM model. Right plot:  Same settings and parameters as the left plot but combining with a Euclid-like DR3 future observations.}
\label{fig:three_extens_params}
\end{figure}


We first show in Fig.~\ref{fig:three_extens_params} the Fisher forecasts on our model’s parameters before projecting into the effective dark energy equation of state parameter for the SKA probes alone in the left panel and after combining with Euclid in the right panel. Adding the different parameterisations decrease the constraining power of our probes, especially on the extension parameters. Therefore, we see that without combining with Euclid the close to $\Lambda$CDM case, though with tigher errors than the far from $\Lambda$CDM, could not be distinguished from the former since its confidence contours fall within the 1~$\sigma$. Combining with Euclid improves the detection to 2~$\sigma$. Note that the errors on close to $\Lambda$CDM are smaller because we are then reducing the degeneracies with the cosmological parameters as seen in the different cells of the triplot with respect to the far from $\Lambda$CDM case. However, because the fiducial in the close to $\Lambda$CDM is small, the relative errors turn out to be much bigger as we see in table~\ref{tab:relerr_mg_src} \\ 

\begin{figure}
\centering
\includegraphics[width=0.5\columnwidth]{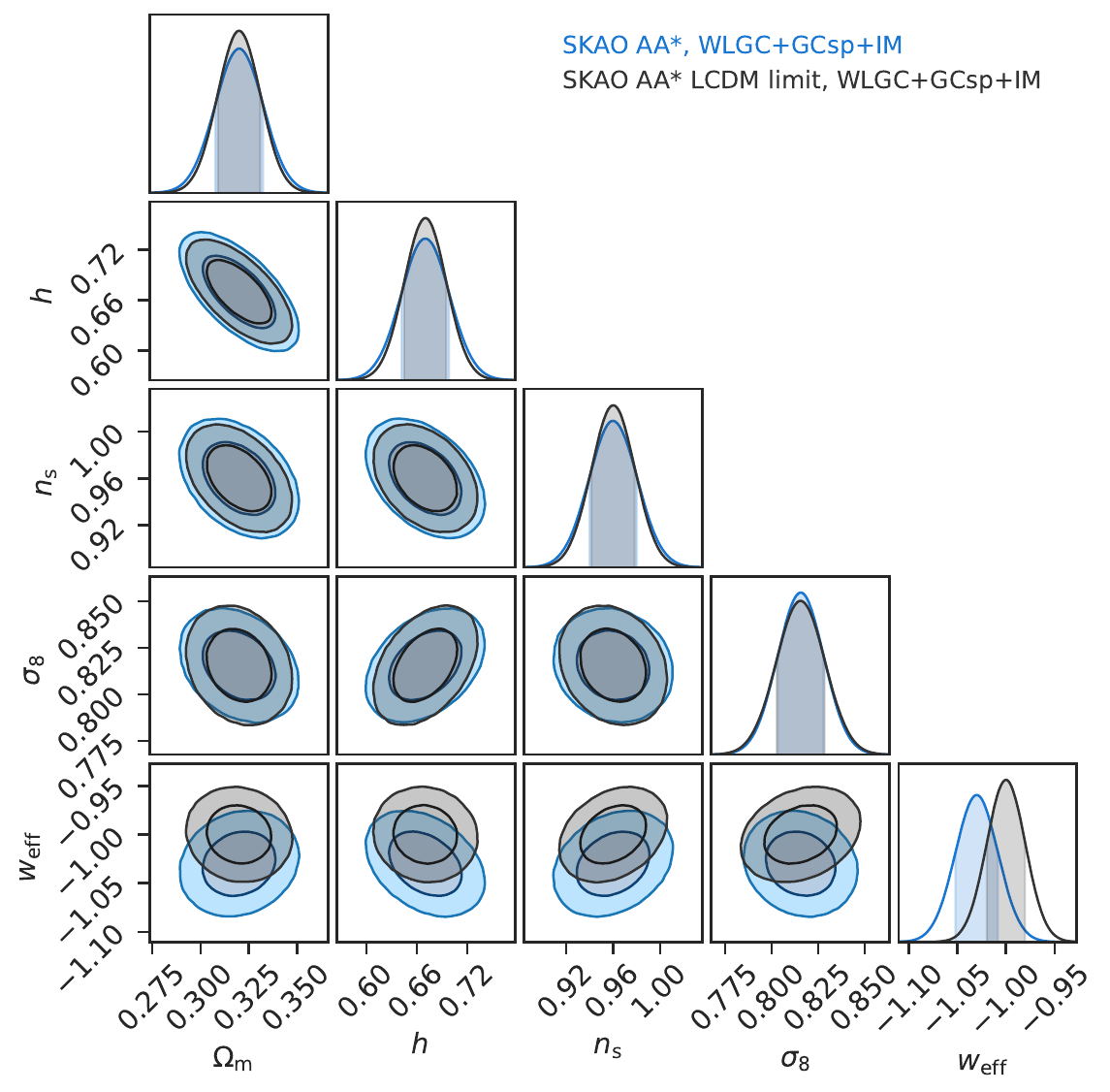}
\includegraphics[width=0.49\columnwidth]{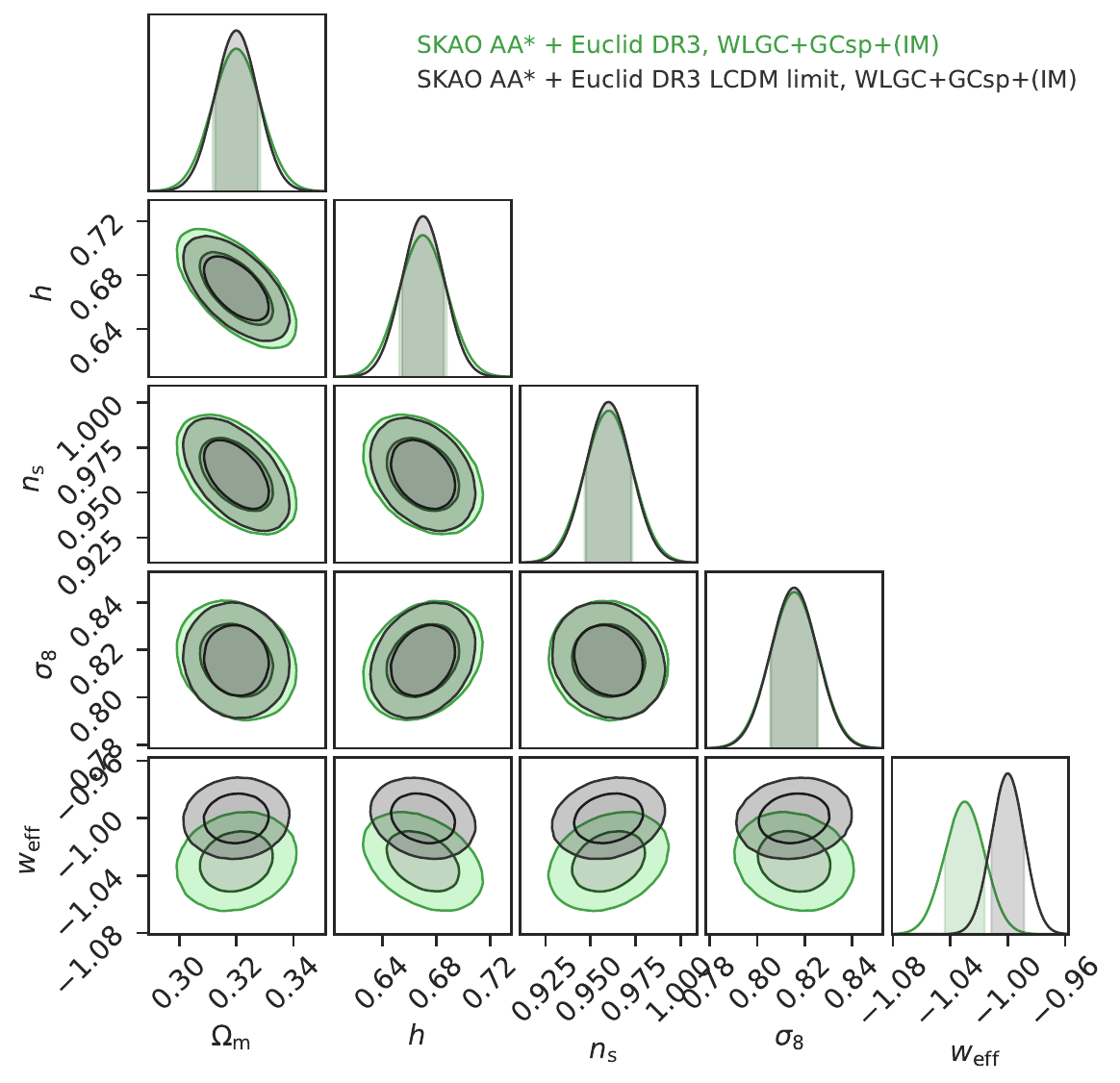}
\caption{Left plot: Forecast 1- and 2-$\sigma$ contours on the cosmological and effective equation of state parameter after projecting the model extension parameters shown in Fig.~\ref{fig:three_extens_params}, using SKAO continuum, HI probes and their combination following the AA* specifications for the far and close to $\Lambda$CDM model. Right plot:  Same settings and parameters as the left plot but combining with a Euclid-like DR3 future observations.}
\label{fig:weff}
\end{figure}

\begin{table}
\centering
\small
\begin{tabular}{lccccccc}
\hline
Case & \texttt{$\Omega_{\rm m}$} (\%)& \texttt{$h$} (\%)& \texttt{$n_{\rm s}$} (\%)& \texttt{$\sigma_8$} (\%)& \texttt{$\mu_{0,\rm mg}$} (\%)& \texttt{$\eta_{0,\rm mg}$} (\%)& \texttt{$H_{0,\rm bck}$} (\%)\\
\hline 
SKAO AA*, & 4.0 & 4.8 & 2.5 & 2.3 & 167 & 262 & 104.4 \\
SKAO AA* $\Lambda$CDM lim  & 3.5 & 3.7 & 2.0 & 1.6& 61332 & 52776 & 41635 \\
SKAO AA* + Euclid DR3,  & 2.7 & 3.0 & 1.6 & 1.5 & 121 & 166 & 67.0 \\
SKAO AA* + Euclid DR3 $\Lambda$CDM lim  & 2.3 & 2.3 & 1.3 & 1.2 & 33933 & 26230 & 23468 \\
\hline
\end{tabular}
\caption{WLGC+GCsp+IM Relative errors  for cosmological and extensions parameters}
\label{tab:relerr_mg_src}
\end{table}
						
We next show in Fig.~\ref{fig:weff} the Fisher forecasts after projecting to obtained constraints on the effective dark energy equation of state parameter for the SKA probes alone in the left panel and after combining with Euclid in the right panel. This in general will tighten the constraints more, after reducing the number of degrees of freedom, but also because of the theoretical modeling relating $w_{\rm eff}$ to the other extension parameters. Therefore we see that the errors on the cosmological parameters are reduced by 30\% and the $w_{\rm eff}$ errors are now comparable to the cosmological ones. We also observe that the errors between the close and far from $\Lambda$CDM are also now comparable since the projection to $w_{\rm eff}$ is based on closer fiducial values for $w_{\rm eff}$ than it was the case for either $\mu_{0,\rm mg}$, $\eta_{0,\rm mg}$ or $H_{0,\rm bck}$ and because we see the same degeneracy level between $w_{\rm eff}$ and the remaining cosmological parameters. However, the reduction is similar for either the close or far from $\Lambda$CDM case so that we end up by obtaining, though with improvement, similar difference in terms of $\sigma$ between the close and far from $\Lambda$CDM $w_{\rm eff}$ models.

Therefore we conclude that the synergy from both surveys benefits to tighten our constraints, however, our highly generalised parameterised model, although impacting at both the background and the perturbation level, remains hard to disentangle from or to rule out $\Lambda$CDM at the stages we performed our forecasts, highlighting the need for the completion of both surveys in order to hope in reaching the aforementioned target.

\begin{table}
\centering
\small
\begin{tabular}{lccccc}
\hline
Case & \texttt{$\Omega_{\rm m}$} (\%)& \texttt{$h$} (\%)& \texttt{$n_{\rm s}$} (\%)& \texttt{$\sigma_8$} (\%)& \texttt{$w_{\rm eff}$} (\%) \\
\hline 
SKAO AA*,  & 3.9 & 4.2 & 2.1 & 1.5 & 2.1 \\
SKAO AA* $\Lambda$CDM limit,  & 3.5 & 3.7 & 1.9 & 1.6 & 1.9 \\
SKAO AA* + Euclid DR3, & 2.6 & 2.6 & 1.4 & 1.2 & 1.3 \\
SKAO AA* + Euclid DR3 $\Lambda$CDM limit, & 2.3 & 2.3 & 1.3 & 1.2 & 1.1 \\
\hline
\end{tabular}
\caption{WLGC+GCsp+IM Relative errors  for cosmological and $w_{\rm eff}$ parameters}
\label{tab:relerr_w_src}
\end{table}

\bibliographystyle{alpha}
\bibliography{main}

\end{document}